\documentclass[useAMS,usenatbib]{mn2e}

\usepackage{graphicx}
\usepackage{url}
\usepackage{equationarray}
\usepackage{amssymb}

\usepackage{times}

\newcommand{\nc}{\newcommand}  
\newcommand\mnras{ Mon. Not. R. Astron. Soc.}%
\newcommand\pasp{ Pub. Astron. Soc. of the Pacific}%
\newcommand\apj{ Astrophys. J.}%
\newcommand\aap{ Astron. \& Astrophys.}%
\newcommand\araa{ Annu. Rev. Astron. Astrophys.}%
\newcommand\aj{ Astron. Journal}%
\newcommand\apjs{ApJS}%
  
\newcommand{\lsun}{L$_{\rm \odot}$}

\newcommand{\av}{$A_{\rm V}$}%
\newcommand\farcsra{\mbox{$.\!\!^{s}$}}%
\nc{\twCO}{$^{12}$CO}  
\nc{\thCO}{$^{13}$CO}  
\nc{\CeiO}{C$^{18}$O}  
\nc{\cmcub}{\mbox{cm$^{-3}$}}  
\nc{\cmsq}{\mbox{cm$^{-2}$}}  
\nc{\Kkms}{\mbox{K~km~s$^{-1}$}}  
\nc{\kms}{\mbox{km~s$^{-1}$}}  
\nc{\ext}{\mbox{T$_{\rm ex}$}}  
\nc{\nhtwo}{\mbox{N(H$_2$)}}
\nc{\Cone}{\mbox{[\ion{C}{i}]$^3$P$_1$--$^3$P$_0$}}  
\nc{\Ctwo}{\mbox{[\ion{C}{i}]$^3$P$_2$--$^3$P$_1$}}  
\nc{\cxo}{CXO J172337.5-373442}
\nc{\cbyco}{\mbox{C/CO}}
\nc\halpha{H$\alpha$}
\nc\hbeta{H$\beta$}
\nc\hgamma{H$\gamma$}
\nc\hdelta{H$\delta$}
\newcommand{\CaI}{Ca {\sc i}}
\newcommand{\FeI}{Fe {\sc i}}
\newcommand{\MgI}{Mg {\sc i}}
\newcommand{\HeII}{He {\sc ii}}
\newcommand\arcdeg{\mbox{$^\circ$}}%

\title[Distance to \cxo]{Optical/IR counterpart to the resolved X-ray jet source \cxo\ and its distance}

\author[B. Mookerjea et al.]
{\parbox{\textwidth}{Bhaswati Mookerjea$^{1}$\thanks{E-mail:
\texttt{bhaswati@tifr.res.in}},
Pietro Parisi$^{2,3}$, 
Sudip Bhattacharyya$^{1}$, 
Nicola Masetti$^{2}$,\\ 
Thomas Kr\"uhler$^{4,5}$ 
and Jochen Greiner$^{4}$ }\vspace{0.4cm}\\
\parbox{\textwidth}{$^{1}$Department of Astronomy and Astrophysics, Tata Institute of
Fundamental Research, Mumbai 400005, India\\
$^{2}$ INAF -- Istituto di Astrofisica Spaziale e Fisica Cosmica di
Bologna, Via Gobetti 101, I-40129 Bologna, Italy\\
$^{3}$ Dipartimento di Astronomia, Universit\'a di Bologna, via Ranzani 1,
I-40129 Bologna, Italy\\
$^{4}$ Max-Planck-Institut f\"ur extraterrestrische Physik, 
Giessenbachstrasse, 85748 Garching, Germany\\
$^{5}$ Universe Cluster, Technische Universit\"{a}t M\"{u}nchen,
Boltzmannstrasse 2, D-85748, Garching, Germany}}

\begin{document}

\date{Accepted \ldots  Received \ldots }

\pagerange{\pageref{firstpage}--\pageref{lastpage}} \pubyear{2002}

\maketitle 

\label{firstpage}

\begin{abstract} 
We present results of observations in the optical to mid-infrared
wavelengths of the X-ray source \cxo, which was serendipitously
discovered in the {\em Chandra} images and was found to have a fully
resolved X-ray jet.  The observations include a combination of
photometry and spectroscopy in the optical using ground-based telescopes
and mid-infrared photometry using {\em Spitzer}. We detect the
optical/IR counterpart of \cxo\ and identify it to be a G9-V star
located at a distance of 334$\pm$60~pc.  Comparable values of the
hydrogen column densities determined independently from the optical/IR
observations and X-ray observations indicate that the optical source is
associated with the X-ray source.  Since the X-ray luminosity can not be
explained in terms of emission from a single G9-V star, it is likely
that \cxo\ is an accreting compact object in a binary system.  Thus,
\cxo\ is the nearest known resolved X-ray jet from a binary system,
which is not a symbiotic star.  Based on the existing X-ray data, the
nature of the compact object can not be confirmed.  However the low
luminosity of the X-ray point source, 7.1$\times10^{30}$\,\lsun,
combined with estimates of the age of the jet and a lack of detection of
bright outburst, suggests that the X-ray jet was launched during extreme
quiescence of the object. The measured low X-ray luminosity of the jet
suggests the likelihood of such jets being more ubiquitous than our
current understanding.
\end{abstract}


\section{Introduction}

Collimated outflows, or jets, are observed from a wide range of
accreting astronomical objects, starting from supermassive black holes
(BHs) in Active Galactic Nuclei (AGNs) to protostars
\citep{BridlePerley1984,MirabelRodriguez1999,fender2006,Kordingetal2008,
Gdeletal2009}.  This ubiquity along with the examples of association
of accretion and jet launching in black hole systems
\citep{greiner1996} and in AGNs \citep{marscher2002} strongly suggest
that the accretion and the jet launching are coupled. Although being a
universal phenomenon this disk-jet coupling is a fundamental topic of
interest in astrophysics, it is still poorly understood.  Jets from
accreting compact stars (stellar mass BHs, neutron stars (NSs), and
even white dwarfs (WDs)) are additionally interesting, because they
(1) are launched from regions of intense gravitational fields; (2) are
useful to probe much shorter timescales in comparison to those of AGN
jets; and (3) can be used to probe the physics of accreting binaries.
For example, although fossil jets have been observed from quiescent
binaries, with luminosities that are several orders of magnitude lower
than their bright-state luminosity \citep{AngeliniWhite2003}, jets
have not been observed to originate due to the quiescent activities.
In binaries, jets in the X-ray wavelengths are known to originate from
synchrotron and/or thermal Bremsstrahlung emissions
\citep{Solerietal2009}. The X-ray jets thus provide essential high
energy information, which is complementary to that extracted from the
more commonly encountered radio jets.  Jets are typically detected
from resolved structures, or from spectral analysis
\citep{Migliarietal2007a,Migliarietal2007b}.  However, resolved X-ray
jets from BH and NS binaries are rare, and the nearest of them is at
least 4 kpc away from the sun \citep{Liuetal2006,Liuetal2007}.  
The nearest X-ray jets observed so far are from the Symbiotic stars
(SSs) RH Cygni \citep{galloway2004} and R Aquarii \citep{kellogg2007},
both located between 200--250\,pc.  We present observations which
suggest that the X-ray jet from \cxo\ is by far (by more than one
order of magnitude) the nearest resolved X-ray jet from an accreting
compact star that is not an SS.

The faint X-ray point source \cxo\ with a prominent jet
($\alpha_{2000}$ = 17$^h$ 23$^m$ 37\farcsra532, $\delta_{2000}$ =
-37\arcdeg 34\arcmin 41\farcs97) was discovered in a {\em Chandra}
image in 2008 \citep{bhattacharyya08}.  However, this image alone was
insufficient to either conclusively decide on the nature of the
source, or to measure its distance.  Here we use a combination of
photometric and spectroscopic observations in the optical and infrared
(IR) in order to identify \cxo, as well as to measure its distance. We
find that this source is a compact star accreting from a G9~V star
located at a distance of 334$\pm$60\,pc.

\section{Data \& Observations}

\subsection{Mid-Infrared photometry using {\em Spitzer}}

The field containing the source CXO J172337.5-373442 was observed by
the two Legacy programs of {\it Spitzer}: The Galactic Legacy Infrared
Mid-Plane Survey Extraordinaire (GLIMPSE) using IRAC at 3.6, 4.5, 5.8
and 8.0~\micron\ and MIPSGAL, a 24 and 70 Micron Survey of the Inner
Galactic Disk with MIPS; all data are available at the NASA/IPAC
Infrared Science Archive (IRSA).  Using accurate photometric analysis
we detect the source clearly in the 4 IRAC bands and do not detect it
with MIPS at 24~\micron. The mean position of the source in the 4 IRAC
bands is $\alpha_{2000}$ = 17$^h$23$^m$37\farcsra5  $\delta_{2000}$
= −37\arcdeg34\arcmin42\farcs2 with an uncertainty of 1\arcsec.
Thus the position of the infrared source is in excellent agreement
with the {\em Chandra} point source discussed by
\citet{bhattacharyya08}.

\subsection{Optical Spectroscopy}

Two 1800-s medium-resolution optical spectra between 3850 and 7200\,\AA\
of \cxo\ were acquired starting at 20:52 UT of 08 August 2009 with the
1.9-m ``Radcliffe" telescope in Sutherland, South Africa.  A
spectrograph equipped with a 266$\times$1798~pixel SITe CCD is mounted
at the Cassegrain focus of this telescope. Using the Grating \#7 and a
slit of 1\farcs8, we obtained a nominal spectral coverage between 3850
and 7200~\AA\ and a dispersion of 2.3~\AA/pix.  

The spectra were optimally extracted \citep{horne1986} using
IRAF\footnote{IRAF is the Image Analysis and Reduction Facility made
available to the astronomical community by the National Optical
Astronomy Observatories, which are operated by AURA, Inc., under
contract with the U. S. National Science Foundation. It is available at
\url{http://iraf.noao.edu/}}  after performing flat-fielding,
bias-subtraction, cosmic ray rejection and background subtraction.  On
each spectrum, wavelength calibration was performed using Cu-Ar lamps,
while flux calibration was done using the spectrophotometric standard
LTT~9239 \citep{hamuy1992}.  We estimate the wavelength calibration
uncertainty to be $\sim$ 0.5 \AA\ using the positions of background
night sky lines. The two spectra were stacked together to increase the
signal-to-noise ratio.

\subsection{7-band Optical and Near-infrared Photometry using {\em
GROND}}

We acquired simultaneous photometric data (Tab.\,1) in seven optical and
near-IR wavebands by observing \cxo\ in $g^\prime r^\prime i^\prime
z^\prime JHK$, using Gamma-Ray Burst Optical/NearInfrared Detector
(GROND, \citep{Greineretal2008}) mounted at the 2.2 m ESO/MPI telescope
at LaSilla observatory (Chile).  The observations were performed on
February 3, 2010.  We converted the $g^\prime r^\prime i^\prime
z^\prime$ photometry to {\em BVRI} magnitudes using the following transformation
equations \citep{lupton05}:

\begin{equationarray}{lcccccc}
B &=& g^\prime &+& 0.3130(g^\prime - r^\prime) &+& 0.2271\\
V &=& g^\prime &-& 0.5784(g^\prime - r^\prime) &-& 0.0038\\
R &=& r^\prime &-& 0.2936(r^\prime - i^\prime) &-& 0.1439\\
I &=& i^\prime &-& 0.3780(i^\prime - z^\prime) &-& 0.3974
\end{equationarray}

\section{Results}

\begin{figure*}
\begin{center}
\includegraphics[angle=-90,width=0.9\textwidth]{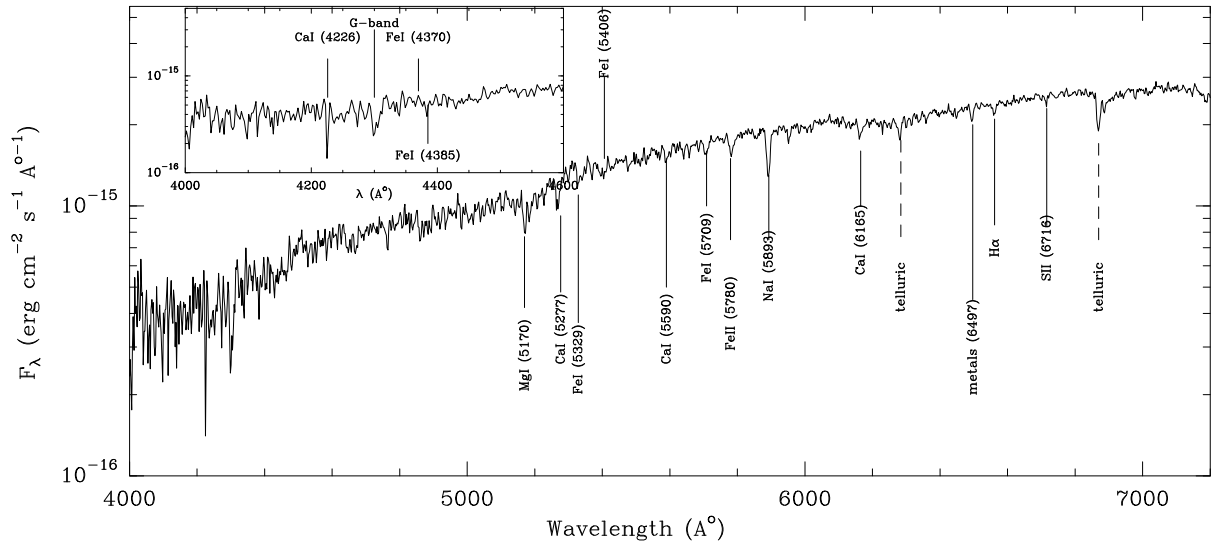}
\caption{The optical spectrum of \cxo\ between 4000 and 7200~\AA. The
inset shows the spectrum between 4000 and 4600~\AA\ in more detail.
Based on the absorption features the source is identified to have a
spectral type of G9~V.  Additional spectral features include the
Diffuse Interstellar Bands (DIBs)  at 5784 and 6284~$\AA$ and a merged
feature due to the Li doublet and combination of multiple weak neutral
iron lines around 6215~\AA.
\label{fig_optspec}}
\end{center}
\end{figure*}

Table~\ref{tab_fluxes} gives the magnitudes/fluxes for \cxo\ in the 
optical to mid-IR wavebands, and Fig.~\ref{fig_optspec} displays the 
optical spectrum between 4000 and 7200~\AA, with some of the main spectral 
features marked on it. We have considered the different possibilities 
regarding the nature of \cxo\ using all the available photometric and 
spectroscopic data.

Based on the MIR photometry obtained with IRAC and MIPS we have obtained 
the following colours for \cxo: [3.6]--[4.5] = -0.26, [5.8]--[8.0] = 0.28,
[3.6]--[5.8] = 0.22 and [8]-[24] = 1.98. Using the observed colours in 
the two diagnostic colour-colour diagrams [3.6]--[4.5] vs [5.8]--[8.0] and
[3.6]--[5.8] vs [8.0]-[24] and comparing the regions occupied by the
stellar and protostellar candidates \citep[cf.][]{mookerjea09} we 
conclude that the source is definitely not a protostellar candidate, but
rather shows colours of a star.

Next, we have searched for the typical optical spectral features seen
in X-ray binaries in our observed spectrum. While \halpha\ is clearly
visible in absorption, \hbeta\, \hgamma\ and \hdelta\ are only
marginally detected, and \HeII\ (4686 \AA) is not detected at all.
Furthermore, the \halpha\ line-shape does not show obvious signs of an
accretion disc. 

Among the several spectral features seen in the optical spectrum,
presence of the spectral lines of \CaI\ (4226 \AA), G-band (4290
\AA), \MgI\ (5170 \AA), \CaI\ (5277 \AA), \FeI\ (5329 \AA), \CaI\
(5590 \AA), \FeI\ (5709 \AA) and \CaI\ (6165 \AA) conclusively proves
the optical counterpart of the source is a G-type star
\citep{hernandez05}.  Other prominent spectral features identified in
the spectrum are marked in Fig.~\ref{fig_optspec}.  The fact that the
Ca I (4226~\AA) line is significantly stronger than the H$\gamma$ and
the H$\delta$ lines further shows that the star is of a later type than G5.
Based on a more precise comparison with stellar spectra from the
library by \citet{jacoby1984} we have identified the source to be a
star of spectral type G9 V.

\begin{figure}
\begin{center}
\includegraphics[angle=0,width=8.0cm,angle=0]{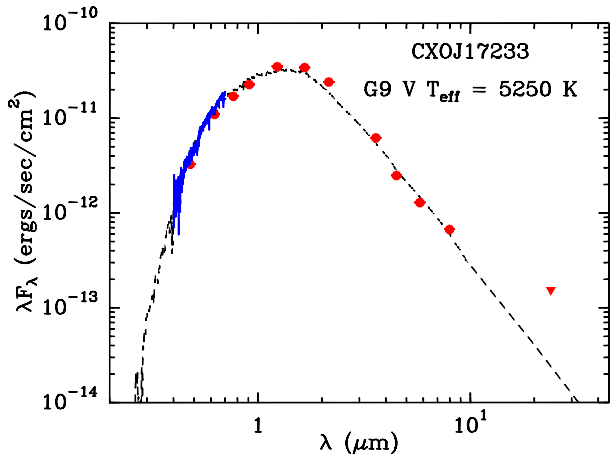}
\caption{Comparison of the emission from a reddened ($A_{\rm V}$ =
2.7) photospheric model of a G9 V type star (with $T_{\rm eff}$ = 5250~K) 
from \citet{kurucz93} (black
dashed line) with the observed optical and IR photometric
fluxes (red points) and optical spectrum (blue continuous line). The
distance to the star is 334~pc.
\label{fig_kuruczsed}}
\end{center}
\end{figure}

The intrinsic color $(B-V)_0$ of a G9 V star is 0.81 \citep{binney98} and
from {\em GROND} the observed color is $(B-V)$=1.66, resulting in
$E(B-V) = 0.85$. We thus obtain a visual extinction of \av $=
2.7\pm0.4$, considering 0.1~mag error in each of the two optical
magnitudes.  This \av value corresponds to a neutral hydrogen column
density ($N_{\rm H}$) of $(5.1\pm0.8)\times 10^{21}$~cm$^{-2}$ (using
the empirical formula of \citet{predehl95}), which is consistent with the
{\em Chandra} X-ray data analysis best-fit value
$3.7^{+1.7}_{-1.4}\times 10^{21}$~cm$^{-2}$ within 90\% confidence
level.  Using the calculated $V$-band magnitude, the estimated \av\ and
the absolute magnitude of a G9 V type star \citep[5.9;][]{binney98}, we
estimate the distance to \cxo\ to be 334$\pm60$~pc. Note that the
comparison of a reddened model stellar atmosphere \citep{kurucz93} for G9
V type star ($T_{\rm eff}$ = 5250~K) with all the observed optical,
near- and mid-IR data (Fig.~\ref{fig_kuruczsed}) shows that such a model
reproduces the observed spectral energy distribution reasonably well,
with the photospheric emission at 24~\micron\ below the detection limit
in the MIPS map.

\begin{center}
\begin{table*}
\caption{Results of optical and IR photometry of \cxo. \label{tab_fluxes}}
\begin{tabular}{rrrrrrrrrrrr}
\hline
\hline
g$^\prime$ & r$^\prime$ & i$^\prime$ & z$^\prime$ & J & H & K & $F_{3.6}$  & F$_{4.5}$ &  F$_{5.8}$ &  F$_{8.0}$ &  F$_{24}$  \\
mag & mag & mag & mag & mag & mag & mag & mJy & mJy & mJy & mJy & mJy\\
\hline
17.1  & 15.5  & 14.8  & 14.3  & 13.5  & 13.2  & 13.3  & 
 7.46  & 3.75  & 2.50  & 1.80 & $<$1.24\\ 
 $\pm$0.1 & $\pm$0.1  &$\pm$0.1 &$\pm$0.1 & $\pm$0.1 &$\pm$0.1 &
 $\pm$0.1 & $\pm$0.02 & $\pm$0.02 & $\pm$0.02 & $\pm$0.03 &\\
\hline
\hline
\end{tabular}
\end{table*}
\end{center}

\section{Discussion}

Here we explore the nature of the compact X-ray source. Since
the IR/optical counterpart of \cxo\ is a G9 V star, this source is not
a protostar or a pulsar wind nebula or a symbiotic star, and is
either a low-mass X-ray binary (LMXB) or a cataclysmic variable (CV).
Note that \citet{bhattacharyya08} excluded the plausible AGN
identification of the source based on the observed low $N_{\rm H}$
value.  

For a Roche lobe filling main-sequence G9 V star, the orbital period
of the system is $\approx 6.3$ hrs (using equation 3.45 of
\citet{BhattacharyavandenHeuvel1991}).  For the estimated $N_{\rm H}$,
an unabsorbed flux of 5.31$\times$10$^{-13}$ erg cm$^{-2}$ s$^{-1}$ in
0.3--10.0 keV range implies a low X-ray luminosity of the point source
($\sim 7.1\times10^{30}$ erg s$^{-1}$; assuming isotropic emission).
In contrast the luminosity of the jet is only
3.9$\times$10$^{29}$\,erg~s$^{-1}$. We note that the expected X-ray
emission of a G9 star is soft and of the order of
10$^{29}$\,erg~s$^{-1}$. Since the expected Eddington luminosity of an
LMXB is $\sim 10^{38-39}$ erg s$^{-1}$, \cxo\ should have been in an
extreme quiescent state during the observation, if this source is an
LMXB. Even if this source is a CV, its low luminosity indicates a
quiescent level.  In this state, the optical and IR emission from the
star can be much more than X-ray emission due to accretion: this is
consistent with the results shown in the Fig.\,4 of
\citet{bhattacharyya08}.  As a result, the signatures of accretion
(e.g., broadened Balmer lines \&  \HeII\ 4686 \AA, all in emission)
may not be discernible in the optical spectrum, which is consistent
with the observation.  The compact star of \cxo\ can thus be a BH or
an NS or a WD.  Although \citet{bhattacharyya08} did not detect a soft
thermal X-ray spectral component from the point source, which is a
typical signature of quiescent NS LMXBs \citep{McClintocketal2004},
such a thermal component cannot be completely ruled out based on the
available {\em Chandra} data. The estimated upper limit of the flux of
the thermal component is about 50\% of the total flux.

Fitting the jet X-ray spectrum with a thermal Bremsstrahlung model
gives an unphysically high best-fit temperature ($\approx 200$ keV).
Therefore, assuming a synchrotron origin of the jet, we follow
\citet{longair1994} and \citet{fender2006} to estimate the minimum
energy associated with \cxo.  The jet length of $\approx 48$\arcsec\
and a width of $\approx 13$\arcsec, roughly give a volume of
$1.0\times10^{51}$\,cm$^{3}$, and hence a minimum jet energy $E_{\rm
min} \geq 1.5\times 10^{41}$\,erg. This minimum energy condition is
achieved when there is an equipartition of energy in particles and the
magnetic field. The corresponding magnetic field is $B_{\rm eq}$ =
28.4~$\mu$G and the Lorentz factor $\gamma$ of the energetic electrons
which emit the synchrotron radiation is 1.6$\times 10^8$. Note that this
is not the Lorentz factor corresponding to the bulk motion of the
material in the jet.

An important question is whether the resolved X-ray jet was ``fossil"
or it originated from quiescent activities. In order to find out, we
assume that the intensity difference between the approaching jet
(observed) and a plausible receding jet (not detected) is solely
because of Doppler boosting and deboosting. Unlike many pulsars the
speeds of binary systems are known to be small, the only two
exceptions being XTE\,J1118+480 and GRO\,J1655-40
\citep{mirabel2001,mirabel2002}.  Hence one of the jets of \cxo\
should not be destroyed by the collision with ISM \citep[as might have
happened for the Geminga pulsar;][]{pavlov2006} and our assumption of
the intensity difference arising only due to Doppler boosting and
deboosting is reasonable.  With this assumption, the photon flux ratio
of the approaching jet to the receding jet is given by
$[(1+\beta\cos\theta)/(1-\beta\cos\theta)]^{2+\Gamma}$, where $\beta
=$ bulk speed of jet in the unit of the speed of light in vacuum,
$\theta =$ angle between the approaching jet's direction and the
observer's direction, and $\Gamma$ is the photon index for a spectral
fitting with a powerlaw. Since the receding jet was not detected
\citep{bhattacharyya08}, we consider the maximum possible limit of
$\beta\cos\theta$, which  is 1. This implies that the upper and lower
limits of $\beta$ and $\theta$ are 1 and $0^{\rm o}$ respectively. In
order to calculate the opposite limits, we use the $3\sigma$ upper
limits of $\Gamma$ ($=2.17$) and the receding jet photon counts
($=9.97$).  This gives a lower limit of $\beta\cos\theta = 0.1$,
implying $\beta > 0.1$ and $\theta < 84^{\rm o}$.  The nearest
detected part of the jet was $\approx 14$\farcs5 away from the point
source during the 4th September 2001 {\em Chandra} observation. Hence,
for $\theta > 11^{\rm o}.5$ (which has 98\% probability of occurrence
assuming $\theta$ could have any random value in $0^{\rm o}-90^{\rm
o}$), and the measured source distance ($= 334$ pc), this nearest
detected part was ejected less than 3.6 years before September 2001.
\cxo\ definitely did not become bright (non-quiescent) after January
1998, because, otherwise the {\em Rossi X-ray Timing Explorer} ({\em
RXTE}) ``All Sky Monitor" and {\em RXTE} ``Proportional Counter Array"
Galactic bulge scan would have detected it. The X-ray jet of this
source was most likely launched by the quiescent activities.
Furthermore, the ratio of the jet power to the (quiescent) accretion
power is $\approx 0.055$ for \cxo\ in the $0.3-10$ keV spectral range
(assuming isotropic emission), which is within the same order of
magnitude of the typical ratio ($\approx 0.1$) for a range of
astrophysical objects \citep{Kordingetal2008}. 
	

\section{Summary}

We have identified the optical/IR counterpart of the X-ray
source \cxo\ as a G9-V star and determined the distance to the source
to be 334$\pm$60\,pc.  Based on the comparable values of hydrogen
column densities measured independently toward the optical/IR source
and the X-ray source, we conclude that these two most likely form a
binary.  Barring the two Symbiotic stars RH Cygni and R Aquarii no
other binary system with resolved X-ray jet has been observed at
such short distances from us. The currently available X-ray data does
not allow us to determine the nature of the compact object. In the
event of \cxo\ being an BH LMXB, it would the nearest known BH. The
low X-ray luminosity of the point source, 7.1$\times
10^{30}$\,erg\,s$^{-1}$, suggests that the jet is most likely launched
during quiescent activity. Further, the low luminosity of the jet,
$3.9\times10^{29}$ erg s$^{-1}$, implies that such jets could be
significantly more ubiquitous than are currently known to be, since
being located at large distances they are not likely to be detected
easily.

\section{acknowledgements}
 
We thank S. Klose, A. Nicuesa (both Tautenburg Observatory) and F.
Schrey (MPE Garching) for support during the GROND observations.  NM
and PP acknowledge the ASI and INAF financial support via grant No.
I/008/07. PP is supported by the ASI-INTEGRAL grant No. I/008/07.   NM
thanks Domitilla de Martino for useful discussions on jets in CVs.
Part of the funding for GROND (both hardware as well as personnel) was
generously granted from the Leibniz-Prize to Prof. G. Hasinger (DFG
grant HA 1850/28-1). This research has made use of the ASI Science
Data Center Multimission Archive; it also used the NASA Astrophysics
Data System Abstract Service, and the NASA/IPAC Infrared Science
Archive, which are operated by the Jet Propulsion Laboratory,
California Institute of Technology, under contract with the National
Aeronautics and Space Administration.  This publication made use of
data products from the Two Micron All Sky Survey (2MASS), which is a
joint project of the University of Massachusetts and the Infrared
Processing and Analysis Center/California Institute of Technology,
funded by the National Aeronautics and Space Administration and the
National Science Foundation.

\label{lastpage}


\begin{thebibliography}{}

\bibitem[\protect\citeauthoryear{Angelini \&
White}{2003}]{AngeliniWhite2003} Angelini L., White N.~E., 2003, ApJ,
586, L71 

\bibitem[Bhattacharya \& van den Heuvel (1991)]{BhattacharyavandenHeuvel1991}
Bhattacharya, D., \& van den Heuvel, E. P. J. 1991, Physics Reports, 203, 1

\bibitem[Bhattacharyya(2008)]{bhattacharyya08} Bhattacharyya, S.\ 2008, 
\mnras, 391, L117 

\bibitem[Binney \& Merrifield(1998)]{binney98} Binney, J., \& Merrifield, M.\ 1998, Galactic astronomy : Princeton University Press

\bibitem[Bridle \& Perley(1984)]{BridlePerley1984} A. H. Bridle, R. A.
Perley, 1984, \araa, 22, 319 

\bibitem[\protect\citeauthoryear{Eikenberry et 
al.}{2008}]{eikenberry2008} Eikenberry S.~S., Patel S.~G., Rothstein 
D.~M., Remillard R., Pooley G.~G., Morgan E.~H., 2008, ApJ, 678, 369 

\bibitem[\protect\citeauthoryear{Fender}{2006}]{fender2006} Fender 
R., 2006, csxs.book, 381 

\bibitem[\protect\citeauthoryear{Galloway \&
Sokoloski}{2004}]{galloway2004} Galloway D.~K., Sokoloski J.~L., 2004,
ApJ, 613, L61 


\bibitem[\protect\citeauthoryear{Greiner, Morgan, 
\& Remillard}{1996}]{greiner1996} Greiner J., Morgan E.~H., Remillard R.~A., 1996, ApJ, 473, L107 

\bibitem[\protect\citeauthoryear{Greiner et 
al.}{2008}]{Greineretal2008} Greiner J., et al., 2008, PASP, 120, 405 

\bibitem[\protect\citeauthoryear{G{\"u}del et 
al.}{2009}]{Gdeletal2009} G{\"u}del M., Skinner S.~L., Cabrit S., 
Eisl{\"o}ffel J., Dougados C., Gredel R., Briggs K.~R., 2009, pjc..book, 
347 

\bibitem[Hamuy et al.(1992)]{hamuy1992} Hamuy, M., Walker, A.~R., 
Suntzeff, N.~B., Gigoux, P., Heathcote, S.~R., 
\& Phillips, M.~M.\ 1992, \pasp, 104, 533 

\bibitem[Hamuy et al.(1994)]{hamuy1994} Hamuy, M., Suntzeff, 
N.~B., Heathcote, S.~R., Walker, A.~R., Gigoux, P., 
\& Phillips, M.~M.\ 1994, \pasp, 106, 566 

\bibitem[Hern{\'a}ndez et al.(2005)]{hernandez05} Hern{\'a}ndez, 
J., Calvet, N., Hartmann, L., Brice{\~n}o, C., Sicilia-Aguilar, A., 
\& Berlind, P.\ 2005, \aj, 129, 856 

\bibitem[Horne(1986)]{horne1986} Horne, K.\ 1986, \pasp, 98, 609 

\bibitem[Jacoby et al.(1984)]{jacoby1984} Jacoby, G.~H., Hunter, 
D.~A., \& Christian, C.~A.\ 1984, \apjs, 56, 257 

\bibitem[\protect\citeauthoryear{Kellogg et 
al.}{2007}]{kellogg2007} Kellogg E., Anderson C., Korreck K., 
DePasquale J., Nichols J., Sokoloski J.~L., Krauss M., Pedelty J., 2007, 
ApJ, 664, 1079 


\bibitem[\protect\citeauthoryear{K{\"o}rding et 
al.}{2008}]{Kordingetal2008} K{\"o}rding E., Rupen M., Knigge C., 
Fender R., Dhawan V., Templeton M., Muxlow T., 2008, Sci, 320, 1318 

\bibitem[Kurucz(1993)]{kurucz93} Kurucz, R. 1993, Kurucz CD-ROM No. 13, 
ATLAS9 Stellar Atmosphere Programs and 2 km/s Grid (Cambridge: SAO)

\bibitem[\protect\citeauthoryear{Liu, van Paradijs, \& van den
Heuvel}{2006}]{Liuetal2006} Liu Q.~Z., van Paradijs J., van den Heuvel
E.~P.~J., 2006, A\&A, 455, 1165 

\bibitem[\protect\citeauthoryear{Liu, van Paradijs, \& van den
Heuvel}{2007}]{Liuetal2007} Liu Q.~Z., van Paradijs J., van den Heuvel
E.~P.~J., 2007, A\&A, 469, 807 

\bibitem[\protect\citeauthoryear{Longair}{1994}]{longair1994} 
Longair M.~S., 1994, High 
energy astrophysics.~Volume 2.~Stars, the Galaxy and the interstellar 
medium., Cambridge University Press, Cambridge (UK)

\bibitem[Lupton(2005)]{lupton05}R. Lupton
\url{http://www.sdss.org/dr7/algorithms/sdssUBVRITransform.html#Lupton2005}

\bibitem[\protect\citeauthoryear{Marscher et 
al.}{2002}]{marscher2002} Marscher A.~P., Jorstad S.~G., G{\'o}mez 
J.-L., Aller M.~F., Ter{\"a}sranta H., Lister M.~L., Stirling A.~M., 2002, 
Natur, 417, 625 

\bibitem[McClintock et al. (2004)]{McClintocketal2004}
McClintock, J. E., Narayan, R., \& Rybicki, G. B. 2004, \apj, 615, 402

\bibitem[\protect\citeauthoryear{Migliari et 
al.}{2007a}]{Migliarietal2007a} Migliari S., et al., 2007, \apj, 670, 610 

\bibitem[\protect\citeauthoryear{Migliari et 
al.}{2007b}]{Migliarietal2007b} Migliari S., et al., 2007, \apj, 671, 706 

\bibitem[Mirabel \& Rodr\'iguez(1999)]{MirabelRodriguez1999} I. F.
Mirabel, L. F. Rodr\'iguez, \araa 37, 409 (1999)

\bibitem[\protect\citeauthoryear{Mirabel et 
al.}{2001}]{mirabel2001} Mirabel I.~F., Dhawan V., Mignani R.~P., 
Rodrigues I., Guglielmetti F., 2001, Natur, 413, 139 

\bibitem[\protect\citeauthoryear{Mirabel et 
al.}{2002}]{mirabel2002} Mirabel I.~F., Mignani R., Rodrigues I., Combi J.~A., Rodr{\'{\i}}guez L.~F., Guglielmetti F., 2002, A\&A, 395, 595 

\bibitem[Mookerjea et al. (2009)]{mookerjea09} Mookerjea, B., Sandell,
G., Jarrett, T.~H., \& McMullin, J.~P.\ 2009, \aap, 507, 1485

\bibitem[Narayan \& McClintock (2008)]{NarayanMcClintock2008}
Narayan, R., \& McClintock, J. E. 2008, New Astronomy Reviews, 51, 733

\bibitem[\protect\citeauthoryear{Pavlov, Sanwal, 
\& Zavlin}{2006}]{pavlov2006} Pavlov G.~G., Sanwal D., Zavlin V.~E., 2006, ApJ, 643, 1146 



\bibitem[Predehl \& Schmitt (1995)]{predehl95} Predehl, P., \& Schmitt, 
J.~H.~M.~M.\ 1995, A\&A, 293, 889

\bibitem[\protect\citeauthoryear{Soleri et al.}{2009}]{Solerietal2009} 
Soleri P., Tudose V., Fender R., van der Klis M., Jonker P.~G., 2009, 
MNRAS, 399, 453

\bibitem[\protect\citeauthoryear{Skrutskie et 
al.}{2006}]{skrutskie06} Skrutskie M.~F., et al., 2006, AJ, 131, 
1163 

\bibitem[Warner(1995)]{warner1995} Warner, B.\ 1995, Cambridge 
Astrophysics Series, Cambridge, New York: Cambridge University Press, 
|c1995,  


\bibitem[Zacharias et al. (2005)]{zacharias05} Zacharias, N., Monet, 
D.~G., Levine, S.~E., Urban, S.~E., Gaume, R., \& Wycoff, G.~L.\ 2005, 
BAAS, 36, 1418



\end{thebibliography}
\end{document}